\documentclass[apj,iop, graphicx]{emulateapj}
\usepackage{apjfonts}

\DeclareTextFontCommand{\textmyfont}{\myfont}

\makeatletter
\newcommand*{\rom}[1]{\expandafter\@slowromancap\romannumeral #1@}

\newcommand*{\myfont}{\fontfamily{cmtt}\selectfont}

\begin{document}

\title{Spectral--timing analysis of the lower kHz QPO in the low-mass X-ray binary Aquila X-1}
\shortauthors{Troyer \& Cackett}
\shorttitle{Spectral--timing analysis of Aql X-1}

\author{Jon~S.~Troyer\altaffilmark{1}, Edward~M.~Cackett\altaffilmark{1}}

\email{jon.troyer@wayne.edu}

\affil{\altaffilmark{1}Department of Physics \& Astronomy, Wayne State University, 666 W. Hancock St, Detroit, MI 48201, USA}

\begin{abstract}
\noindent Spectral--timing products of kilohertz quasi-periodic oscillations (kHz QPOs) in low--mass X-ray binary (LMXB) systems, including energy- and frequency-dependent lags, have been analyzed previously in 4U 1608-52, 4U 1636-53, and 4U 1728-34. Here, we study the spectral--timing properties of the lower kHz QPO of the neutron star LMXB Aquila X-1 for the first time.  We compute broadband energy lags, as well as energy-dependent lags and the covariance spectrum using data from the {\it Rossi X-ray Timing Explorer} (RXTE).  We find very similar characteristics to other previously studied systems, including soft lags of  $\sim$30 $\mu$s between the 3.0 -- 8.0 keV and 8.0 -- 20.0 keV energy bands at the average QPO frequency.  We also find lags that show a nearly monotonic trend with energy, with the highest energy photons arriving first.  The covariance spectrum of the lower kHz QPO is well fit by a thermal Comptonization model, though we find a higher seed photon temperature compared to the mean spectrum, which was also seen in \citet{Peille_15}, and indicates the possibility of a composite boundary layer emitting region.   Lastly, we see in one set of observations, an Fe K component in the covariance spectrum at 2.4-$\sigma$ confidence which may raise questions about the role of reverberation in the production of lags.

\end{abstract}

\section{Introduction}
Accretion of matter onto compact objects (black holes and neutron stars) offers an avenue to study the effects of strong gravity as well as potentially constrain the mass and size-scale of these ultra-dense objects.  Accreting neutron stars and black holes occur in binary systems in which the companion acts as a matter donor.  In low-mass X-ray binary systems (LMXB) -- systems where the companion star has a mass $\leq$1 $M_{\odot}$ -- the companion star overflows its Roche lobe and matter is transferred from the companion to the compact object via accretion.  See \citet{vanderKlis_00} for a more detailed overview of accretion and oscillations in LMXB systems.  The distance scale of the inner accretion flow is expected to be on the order of the neutron star radius.  This implies dynamical velocities and timescales of the order of $\simeq 0.5 $c and $\simeq 100 \mu s$ respectively \citep{vanderKlis_00, Wagoner_03}.  We therefore expect signals that carry the causal signatures of this region to have the same timescale.  The shortest timescale (highest frequency) oscillations observed are the kilohetrz quasi-periodic oscillations (kHz QPOs). 

kHz QPOs were discovered shortly after the launch of NASA's {\it Rossi X-ray Timing Explorer} (RXTE) \citep{Bradt_93} in December 1995.  See \citet{vanderKlis_98} for a history of the early days of RXTE's discoveries of kHz QPOs.  The discovery of two distinct kHz QPOs in nearly every neutron star LMXB system containing QPOs led to twin kHz QPO becoming a signature of neutron star systems \citep{vanderKlis_06}.  The QPOs occur in the 300 - 1200 Hz range and were quickly thought to be associated with orbital frequencies of the inner accretion flow - a characteristic shared by a majority of the models that attempt to explain the origin of kHz QPOs \citep{miller98,stella99,lamb01}.  However, there are models that do not associate the kHz QPOs with the orbital frequencies of the inner accretion flow \citep[see e.g.,][]{Lee_01,Kumar_Misra_14,Kumar_Misra_16}. See \citet{vanderKlis_00, vanderKlis_06} for a review of various kHz QPO models.  

Since kHz QPOs occur on timescales of the inner accretion flow of neutron star LMXB systems, we wish to apply spectral-timing techniques in order to probe the geometry of these systems.  See \citet{Nowak_99} and \citet{Uttley_14} for detailed reviews of spectral-timing analysis techniques.  The first energy-dependent soft lags of a neutron star LMXB (4U 1608-52) were found in \citet{Vaughan_98}.   Soft lags occur when the higher energy photons associated with a correlated variation in flux arrive before the lower energy photons.  Additionally, soft lags were also found in other neutron star LMXB systems \citep{Kaaret_99, Barret_13, deAvellar_13,Peille_15}, black hole binaries and AGN \citep[see][for a review of reverberation in black hole systems]{Uttley_14}.  

While \citet{Vaughan_98} and \citet{Kaaret_99} were the first works to study soft time lags in kHz QPOs in 4U 1608-52 and 4U 1636-53 respectively, more recent analyses have been done for a total of three neutron star LMXBs.  Soft lags of kHz QPOs have been studied in 4U 1608-52 in \citet{deAvellar_13} and \citet{Barret_13}, in 4U 1636-53 in \citet{deAvellar_13, deAvellar16}, and 4U 1728-34 in \citet{Peille_15}.  These studies have all shown for the lower kHz QPO:  soft broadband lags and a near monotonic trend of lag with energy, with the higher energy photons arriving first.  The magnitudes of the soft broadband lags have all been on the order of the size scale of the neutron star inner accretion disk/boundary layer.  

The additional spectral analysis done in \citet{Peille_15} for 4U 1608-52 and 4U 1728-34 shows: a harder covariance Comptonization component compared with the time-averaged Comptonization component, as well as a better spectral fit when the seed photon temperature of these two components are decoupled.  For that analysis, the covariance seed photon temperature was found to be systematically higher than the mean spectrum Comptonization component.

In this paper we apply spectral-timing analysis techniques to Aql X-1 with data from {\it RXTE/PCA}.  We discuss our analysis approach, data reduction, and the various data products is Section \ref{data_analysis}.  In Section \ref{discussion} we note similarities between our results and results of previous studies of other neutron star LMXB systems and review some of their implications.  Finally, in Section \ref{conclusion}, we summarize the most important results.

\section{Data Analysis} \label{data_analysis}
\subsection{Overview}
We searched the entire {\it RXTE/PCA} archive for observations of Aql X-1 in modes compatible with spectral-timing analysis.  In all cases, we required better than 128 $\mu$s timing resolution and 64 energy channels.  Once such observations were identified, we required significantly detected kHz QPOs in order to obtain sufficient statistics for meaningful analysis.  Using \citet{Barret_08}, we were able to select observations with significantly detected QPOs up to July 2007.\footnote{ We searched all mode compatible observations after July 2007.  There was a single OBSID (94076-01-05-00) with a single observation where the lower kHz QPO was significantly detected.  However, due to the short duration (2.3 ks) of this observation, we could not produce any spectral-timing products because of the limited statistics.}  It should be noted that in the case of Aql X-1, only a single kHz QPO - likely the lower kHz QPO \citep{Mendez_01} - is detected well enough to perform spectral-timing analysis \citep{Barret_08}.  Following \citet{Barret_13}, we evaluated kHz QPOs by computing the power spectral density (PSD) for each time bin of the lightcurve.  We used a binning time of 256 s, ensuring the bins did not cross individual observations.  We computed the discrete Fourier transform, calculated the periodogram \citep{Uttley_14}, and left it in counts units.  We then searched the PSD for power excess and used the $\chi^2$ method to fit a constant plus a Lorentzian with three parameters: centroid frequency ($\nu$), full-width half maximum frequency ($\Delta$$\nu$), and normalization ($I_{lor}$).  Thus, we obtained a single QPO frequency for each 256 s bin.  A QPO is considered significant if the ratio $I_{lor}/\Delta I_{lor} \ge$ 3.0.

For observations with significantly detected QPOs, there are several ways of presenting the data.  The first is by combining observations within a single OBSID.  For {\it RXTE}, an OBSID is a grouping of observations within a single, contiguous pointing.  In this case there are no issues of changing source state or instrument response since the time intervals between exposures are much shorter than the observation times.  Problems arise however in obtaining sufficient S/N to obtain meaningful results.  In order to expand our analysis, the approach we take is to combine observations in which the instrument response does not vary significantly.  Since the spectral properties of the source itself can change between observations, what we present is an average over the times selected.  The criteria we used to choose how to combine observations was to first verify that the energy channels of interest were the same.  We considered energies from 3.0 keV  to 20.0 keV, above which the background begins to dominate.  Even with the same energy channels, between observations the energy ranges in each bin fluctuate by small amounts.  It is therefore necessary to rebin in energy so that the energy range fluctuation per bin is much smaller than the energy bin width \citep[see e.g.,][]{Peille_15}.  Within all observation groups, the maximum fractional fluctuation of the centroid energy of a bin is 0.17$\%$ and the maximum fluctuation of an energy bin width is 0.18$\%$.  Overall we present three contiguous observational groupings shown in Tables \ref{tab:Obs_1}, \ref{tab:Obs_2}, and \ref{tab:Obs_3}.  All uncertainties throughout the paper are quoted at the 1$\sigma$ level.

\begin{deluxetable}{c c c c c}
\tablewidth{0pt}
\tablecolumns{4}
\tablecaption{Aql X-1 Observation Group 1: Observation Properties}
\tablehead{ObsID  & Date & Event Mode & Exposure & Significant\\
 & mm/dd/yyyy & Counts & Time (s) & QPOs}
\startdata
20092-01-01-02 & 08/13/1997 & 1434551 & 911 & 3\\ 
20092-01-02-01 & 08/15/1997 & 2378037 & 1391 & 1\\ 
20092-01-02-03 & 08/17/1997 & 1470468 & 833 & 3\\ 
20092-01-05-01 & 09/06/1997 &  22695778 & 14263 & 3\\ 
20098-03-07-00 & 02/27/1997 & 5888675 & 4538 & 14\\ 
20098-03-08-00 & 03/01/1997 & 5793703 & 5776 & 13\\ 
30072-01-01-01 & 03/03/1998 & 2498232 & 1393 & 5\\ 
30072-01-01-02 & 03/04/1998 & 3310253 & 1510 & 4\\ 
30072-01-01-03 & 03/05/1998 & 3168559 & 1314 & 6
\enddata
\label{tab:Obs_1}	
\end{deluxetable}

\begin{deluxetable}{c c c c c}
\tablewidth{0pt}
\tablecolumns{4}
\tablecaption{Aql X-1 Observation Group 2: Observation Properties}
\tablehead{ObsID  & Date & Event Mode & Exposure & Significant \\
 & mm/dd/yyyy & Counts & Time (s) & QPOs}
 \startdata
40047-02-05-00 & 05/31/1999 & 13061454 & 9456 & 2\\ 
40047-03-02-00 & 06/03/1999 & 13043680 & 10777 & 4\\ 
40047-03-03-00 & 06/04/1999 & 12172425  & 9831 & 16
\enddata
\label{tab:Obs_2}	
\end{deluxetable}

\begin{deluxetable}{c c c c c}
\tablewidth{0pt}
\tablecolumns{4}
\tablecaption{Aql X-1 Observation Group 3: Observation Properties}
\tablehead{ObsID  & Date & Event Mode & Exposure & Significant \\
 & mm/dd/yyyy & Counts & Time (s) & QPOs}
 \startdata
50049-02-13-00 & 11/07/2000 & 5828947 & 3011 & 2\\ 
50049-02-15-03 & 11/13/2000 & 7268319 & 5456 & 14\\ 
50049-02-15-04 & 11/14/2000 & 4918301 & 5034 & 9\\ 
50049-02-15-05 & 11/15/2000 & 9864954  & 9747 & 1\\ 
50049-02-15-06 & 11/16/2000 & 1807040 & 1949 & 5\\ 
70069-03-01-01 & 03/07/2002 & 2727478 & 2429 & 6\\ 
70069-03-01-02 & 03/07/2002 & 1836713 & 1647& 3\\
70069-03-02-01 & 03/10/2002 & 1460966 & 813 & 4\\ 
70069-03-03-06 & 03/18/2002 & 918008 & 918 & 2\\ 
70069-03-03-07 & 03/18/2002 & 3268159 & 3264 & 4\\ 
70069-03-03-09 & 03/19/2002 & 1388293 & 1288 & 3\\ 
70069-03-03-14 & 03/21/2002 & 2092049 & 2690 & 2
\enddata
\label{tab:Obs_3}	
\end{deluxetable}

\subsection{Data Reduction}
To produce the spectral-timing products, we use the {\it RXTE/PCA} event mode data listed in Tables \ref{tab:Obs_1}, \ref{tab:Obs_2}, and \ref{tab:Obs_3}.  First, in order to determine the conversion from channel to energy, we extract spectra and create associated response matrices using \textmyfont{seextrct} and \textmyfont{pcarsp}.  We applied good time intervals (GTI) to account for PCUs turning on and off, Earth limb avoidance, and avoidance of the South Atlantic Anomaly (SAA).  From the response matrices we get the energy range associated with each binned channel, and determine the absolute channel values using \textmyfont{chantrans}.

For each observation group, we analyzed all event mode files and computed the fast Fourier transform (FFT) at 4.0 seconds (s) intervals which are then averaged over 256 s bins.  Data gaps in the GTIs are windowed and the averaged FFTs are not permitted to cross observations.  Each 256 s bin was then searched for excess power and fit with a 3-parameter Lorentzian as described above.  We discarded any QPOs with significance < 3.0.  Any bursts were not included in our analysis. 

\subsection{Lags vs. Frequency}
\begin{figure}
\centering
\includegraphics[trim=6cm 1.4cm 6cm 1cm, clip,width=8cm]{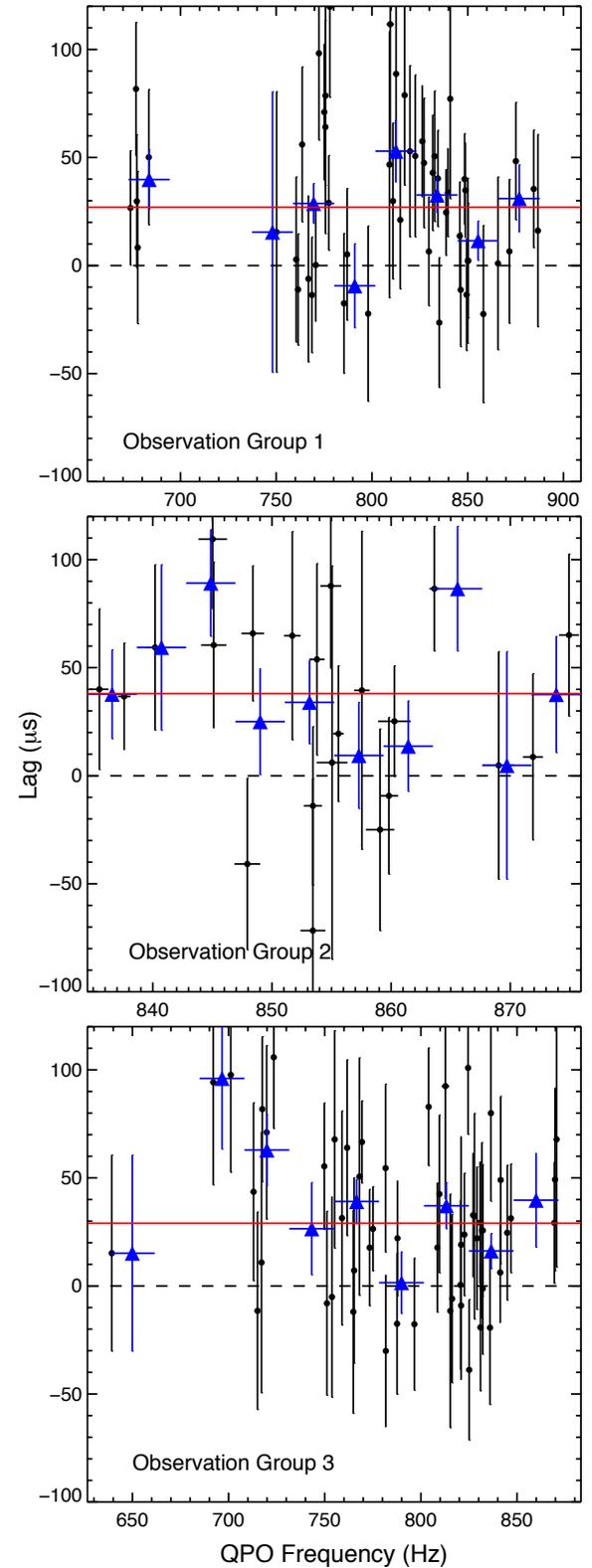}
\caption{Frequency-dependent lags for each observation group. Each data point (small black dot) is the lag from a 256 s bin with a significantly detected kHz QPO.  The mean lag between the 3.0 - 8.0 keV and 8.0 - 20.0 keV bands are shown in red.  Additionally, the lags are binned into 10 equally-spaced frequency bins (blue triangles) between the minimum and maximum QPO frequency. }
\label{fig:Lag_freq}
\end{figure}

\begin{figure}
\centering
\includegraphics[trim=0.6cm 0.1cm 1cm 1cm, clip,width=8cm]{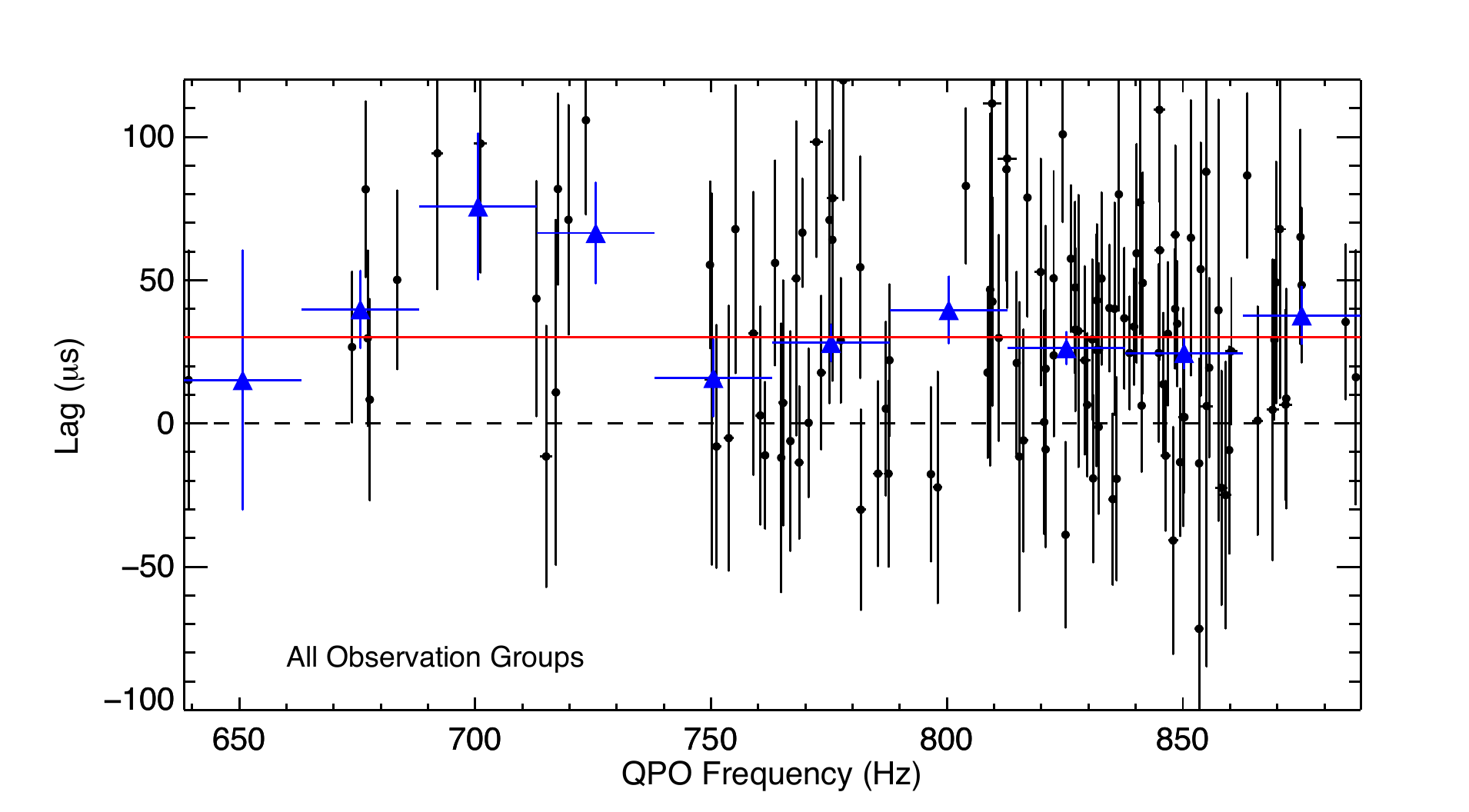}
\caption{Lag as a function of frequency for all observation groups combined. The mean lag is shown in red.  The rebinned data are shown in blue triangles. }
\label{fig:lag_freq_all}
\end{figure}

To establish the presence of any lags, and if there is any frequency dependence, we computed lags between two broad energy bins: 3.0 keV -- 8.0 keV and 8.0 keV -- 20.0 keV.  We computed the cross spectrum for each 256 s data segment between the two energy bins and averaged across the QPO FWHM.  In order to correct for dead time induced cross-talk \citep{vanderKlis_87, Peille_15}, we subtracted Fourier amplitudes between 1350 Hz - 1700 Hz from the cross-spectrum.  We then compute the time lag from the phase of the cross spectrum.   To further characterize the results and highlight any possible trends, we fit a straight line to the data and found fits consistent with no significant dependance of the lag on QPO frequency.  The mean lags for observation group 1, 2, and 3 are $28 \pm 4~\mu$s, $38 \pm 8~\mu$s and $29 \pm 5~\mu$s, respectively, and the mean lag when considering all observations together is $30 \pm 3~\mu$s.  Additionally, we rebinned the lag-frequency data using 10 equally-spaced frequency bins to further illustrate the consistency of lag with frequency.  The lag frequency data for each observation group are shown in Figure \ref{fig:Lag_freq} and the lag frequency data combining observations are show in Figure \ref{fig:lag_freq_all}.  The average lags are all soft lags and positive by convention, indicating that the higher energy band variations lead the lower energy band variations.

\subsection{Lag Energy Spectrum}

\begin{figure}
\centering
\includegraphics[trim=2.5cm 1cm 1.5cm 1cm, clip,width=8cm]{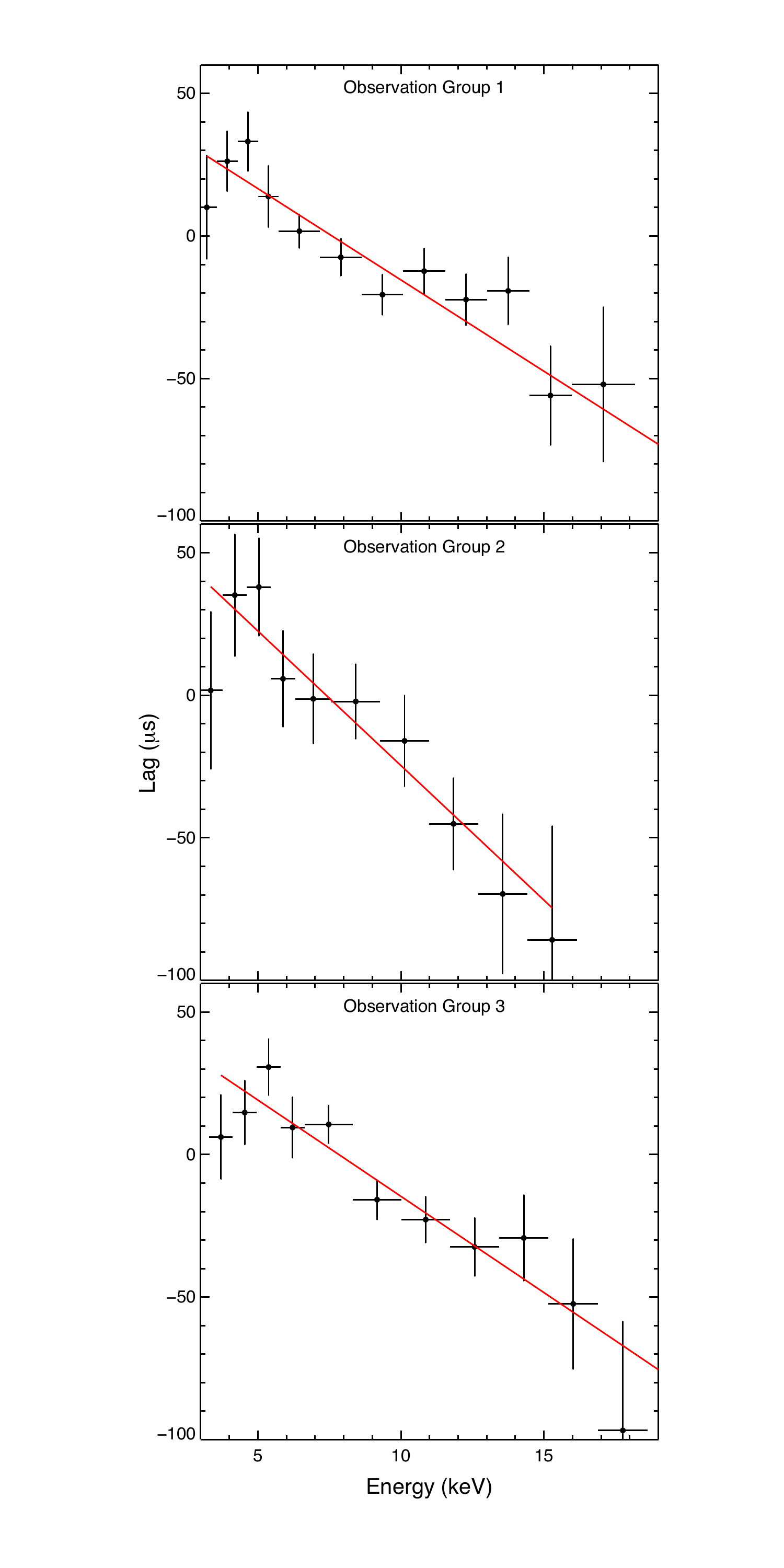}
\caption{Lag-energy spectrum for each observation group.  The lags are computed with respect to a 3.0 keV -- 20.0 keV reference band.  By convention, positive lags indicate photons from that energy bin arrive after the reference band.  Hence the highest energy photons arrive first.  Note that the highest energy bin for observation group 2 could not be calculated due to poor S/N.  The best fit linear relations are also shown.  The best fit parameters are listed in Table \ref{tab:Lagen_fit}.}
\label{fig:Lag_E}
\end{figure}

In order to compute the full lag-energy spectrum, we computed the cross-spectrum within the FWHM of the mean QPO frequency, for each 256 s segment of data,  between each energy band --- channel of interest (CI) --- and the remaining energy channels (3.0 keV - 20.0 keV) --- reference band.  We rebinned in energy, decreasing the number of bins by a factor of 2 in order to increase the signal to noise ratio per bin and to reduce the effect of small energy fluctuations that occur at the channel boundaries between observations mentioned previously.  We then averaged the centroid QPO frequencies and shifted and added \citep{Mendez_98} each cross spectrum to the mean QPO frequency.    We eliminate correlated errors \citep{Uttley_11,Uttley_14} by not including the CI in the reference band.  We then computed the time lag from the phase of mean cross spectrum.  The lag-energy spectra, shown in Figure \ref{fig:Lag_E}, all show nearly monotonic trends with energy, where the highest energy photons arrive before the lower energy photons. We fit each lag-energy spectrum with a straight line to characterize any trend(s).  The data were fit to the function $y = A +Bx$ and are shown in Figure \ref{fig:Lag_E}. The fit parameters are shown in Table \ref{tab:Lagen_fit}.  The best-fitting linear relations are consistent between all 3 observation groups.

\begin{deluxetable}{cccc}
\tablewidth{0pt}
\tablecolumns{3}
\tablecaption{Aql X-1 lag-energy linear fit}
\tablehead{Observation &  A &  B  \\
Group &  ($\mu s$) & ($\mu s$ $keV^{-1}$) }
\startdata
1 &  $49 \pm{8}$ & $-6\pm{1}$ \\ 
2 &  $70 \pm{17}$ & $-9\pm{2}$\\  
3 &  $53 \pm{10}$ & $-7\pm{1}$
\enddata
\tablecomments{Parameters are from the best fit relation $y = A +Bx$}
\label{tab:Lagen_fit}	
\end{deluxetable}

\subsection{Covariance Spectrum}

The covariance spectrum \citep{Wilk_09, Uttley_11} is yet another analysis tool useful in understanding the nature of kHz QPOs and is computed quite easily alongside the lag-energy spectrum.  The equations and methodology for calculating a covariance spectrum are given in detail in \citet{Uttley_14}. The covariance spectrum describes the spectral shape of the portion of the CI which is correlated with the reference band.  Put another way, it is equivalent to the rms spectrum when both are correlated. The first covariance spectrum of a kHz QPO was computed for 4U 1608-52 and 4U 1728-34 in \citet{Peille_15}.  We computed the raw covariance spectrum over the same energy range, and with the same binning and frequencies as the lag-energy spectrum.

In order to compare the covariance spectrum with the time-averaged spectrum, we need to fold the covariance spectrum through the instrument response for the same observation interval. In this way, we can investigate the amount of correlated variability present in each segment of the spectrum.  To get an average instrument response for the covariance spectrum over the observation interval, we expanded the individual response matrices and averaged each entry across observations within a group by weighting it with the fraction of significant QPO time.   

We extracted the Standard 2 spectra for all observations, adding 0.6 \% systematic errors and creating background and response files for each.  We used the most recent bright background model and SAA history.  We verified that the shape of the responses within each observation group were the same (ignoring normalization), with the exception of Observation Group 3, and added the spectra, background and responses.

\begin{figure}
\centering
\includegraphics[trim=0.5cm 8cm 1cm 8cm, clip,width=8cm]{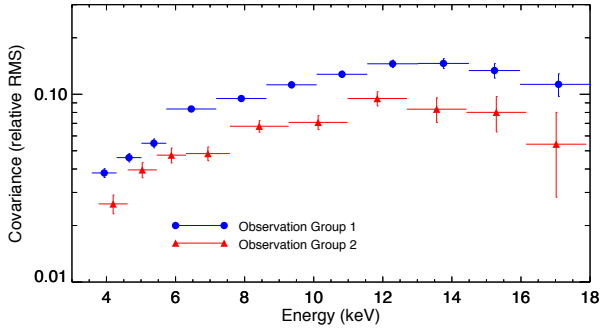}
\caption{Covariance in relative RMS units, which can be thought of the fraction of the spectrum that is variable on the kHz QPO timescale.  The mean spectrum rebinned to match the covariance spectrum binning.  We note an increase in fractional RMS (covariance) with energy up to approximately 12 keV where it levels off.}
\label{fig:rel_rms}
\end{figure}

To calculate the fractional rms (covariance)  we calculate the ratio of the covariance spectrum to the mean spectrum by first rebinning the mean spectrum to match the covariance spectrum binning.  The fractional rms for Aql X-1 is shown in Figure \ref{fig:rel_rms}.  This shows an increase in the fraction of the spectra that is variable with increasing energy, fractional rms (covariance), which becomes nearly constant above approximately 12 keV.  This compares well to previous analyzes of the energy-dependence of the rms in kHz QPOs \citep[e.g.,][]{Mendez_01}.

%\subsubsection{Observation Group 3}
Observation Group 3 showed (3) distinct instrument response profiles in their Standard 2 spectra, which made combining these spectra impossible.  We attempted to break this observation group into (3) corresponding groups, but lack of statistics prevented meaningful calculation of the lag/energy and covariance spectra.  We could therefore not perform any further comparative spectral analysis of Observation Group 3.

\subsection{Spectral Analysis}
We simultaneously fit the mean spectra with the covariance spectra over the 3.0 keV -- 20.0 keV (above 20.0 keV the background dominates) energy range using \textmyfont{XSPEC 12.8.2} \citep{Arnaud_96}.  We use the model combination \textmyfont{phabs*(diskbb+nthcomp+gaussian)} for the fits \citep[see][for a description of \textmyfont{nthcomp}]{Zdz_96, Zycki_99}, though we note that the X-ray spectra of LMXBs are degenerate and can be fit equally well by other model choices \citep[e.g.,][]{Lin_07}.  We fix the photoelectric absorption column density at 0.3 $\times$ 10$^{22}$ cm$^{-2}$ \citep{Kalberla_05}.  For the Fe-line component we use a simple Gaussian model, with centroid constrained between 6.4 keV and 6.97 keV.  Following \citet{Gilf_03, Peille_15} for the covariance spectrum, we use the model combination \textmyfont{phabs*nthcomp} initially with the idea that the covariance spectra might represent the boundary layer emission.

\begin{figure*}
\centering
\includegraphics[trim=1.5cm 0.7cm 0cm 0cm, clip,width=8.5cm]{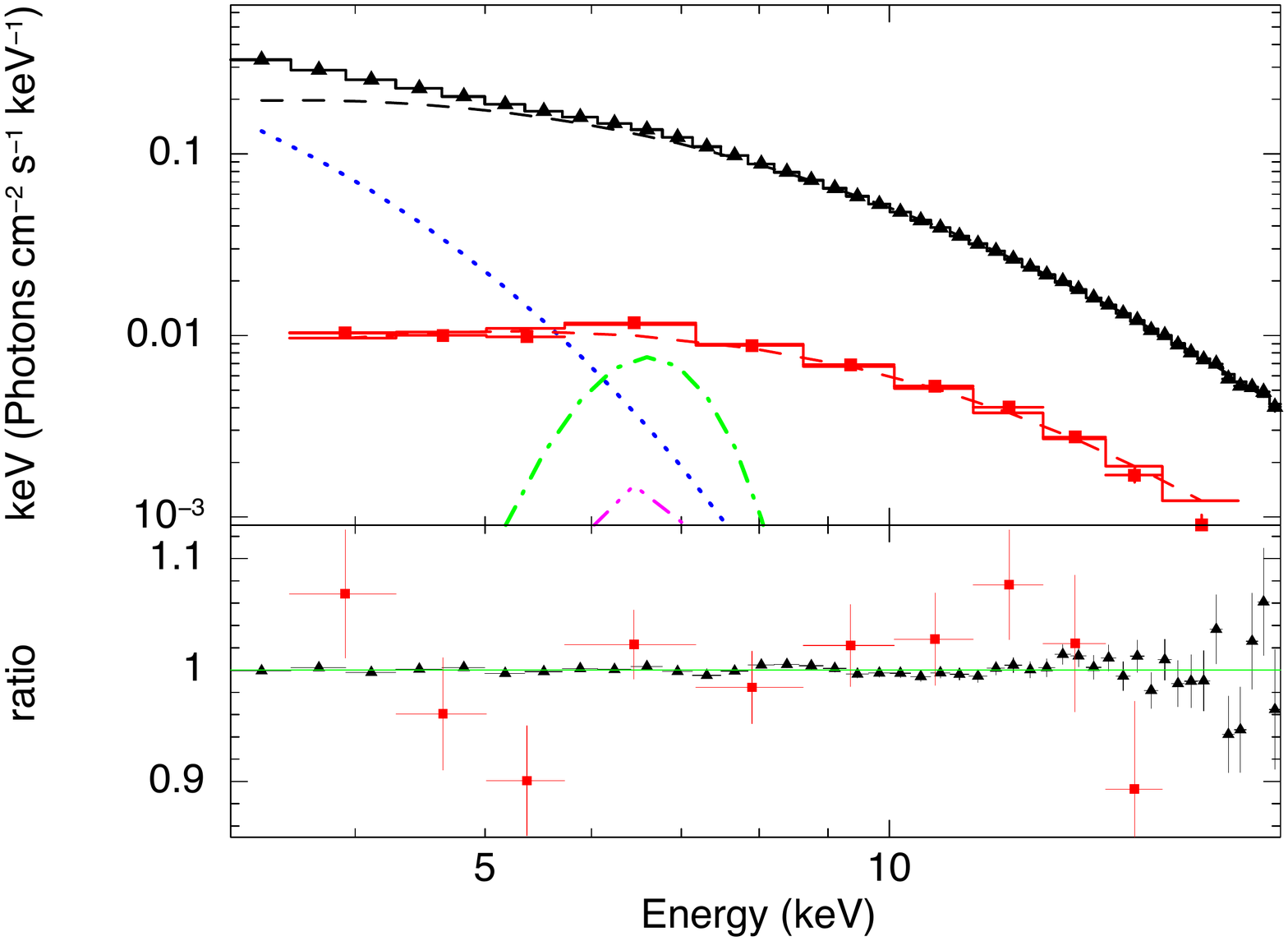}
\hspace{0.5cm}
\includegraphics[trim=1.5cm 0.7cm 0cm 0cm, clip,width=8.5cm]{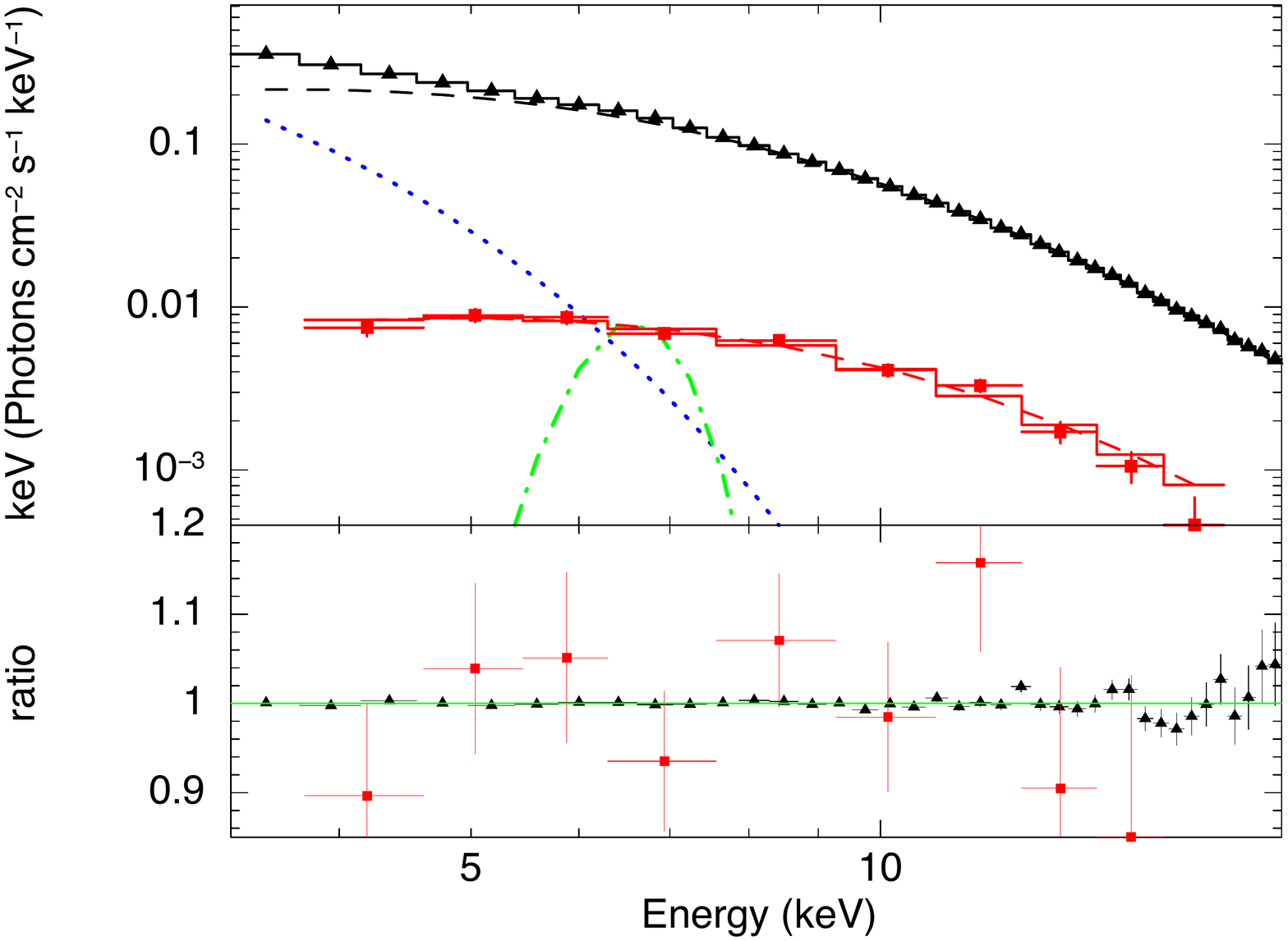}
\caption{{\it Top:} Time-averaged spectrum (black triangles) and covariance spectrum (red squares) for observation group 1 ({\it left}) and observation group 2 ({\it right}). Solid lines indicate the best-fitting overall model. The nthcomp (black dashed), disk blackbody (blue dotted) and Gaussian (green dashed dotted) components for the time-averaged component are shown, while the nthcomp (red dashed) and Gaussian (magenta dashed dotted) components are shown for the covariance spectrum.  There is no Gaussian for the covariance spectrum for observation group 2.  {\it Bottom:} Ratio of the data to the best-fitting model.}
\label{fig:Cov}
\end{figure*}

We find as in \citet{Gilf_03, Peille_15} for 4U 1608-52 and 4U 1728-34, good fits with the chosen model configuration.  We attempted fitting schemes by systematically untying one parameter at a time.  These were: the electron temperature (kT$_e$), photon index ($\Gamma$) and seed photon temperature (kT$_{seed}$).  In order to obtain a good fit, only the seed photon temperature can be untied between the spectra.  All other configurations resulted in poor fits.  We find as in \citet{Peille_15} the seed photon temperature to be systematically higher for the covariance spectrum.  Additionally, in the case of observation group 1, the spectra are fit better when an Fe K Gaussian is included in the covariance spectrum.  In this case, we tied the Gaussian centroid and width of both spectra allowing only the normalizations to vary. With the additional Gaussian in the covariance spectrum, we get a change of $\Delta\chi^2 = 10.1$ for 1 additional degree of freedom, which corresponds to a better fit at the 2.4-$\sigma$ confidence level using an F-test.  In order to further test the presence of the covariance gaussian, we compared fits with no parameters tied between the mean spectrum and covariance spectrum with and without covariance Gaussian.  In this case we also obtain better fits including the covariance Gaussian with a $\Delta\chi^2 = 6.21$ for 1 additional degree of freedom.  This corresponds to a better fit at the 2.0-$\sigma$ confidence level using an F-test.  The spectral decompositions are shown in Figure \ref{fig:Cov} and the best-fitting parameters are listed in Table \ref{tab:fit}.  The model begins to over estimate the covariance spectrum at higher energies.  This is an artifact produced by allowing only a single model parameter to be free for the fits.  This artifact vanishes when both $\Gamma$ and kT$_{seed}$ are freed, with a negligible $\Delta\chi^2$.

Finally, It should be noted that modeling a covariance spectrum with an XSPEC model implicitly assumes that only the normalization is oscillating, but the covariance spectra could also be produced by the average spectrum changing shape, e.g. the seed photon temperature or the optical depth.

\begin{deluxetable*}{lccc}
\tablecaption{Spectral Fit Parameters}
\tablewidth{0pt}
\tablecolumns{4}
\tablehead{
Obs. group & 1* & 1  & 2}
\startdata
N$_{\rm H}$  ($10^{22}$ cm$^{-2}$) & 0.3 (fixed) & 0.3 (fixed) & 0.3 (fixed)\\
kT$_{disk}$(mean) & $0.64\pm0.04$ & $0.65^{+0.07}_{-0.04}$ & $0.67\pm{0.04}$\\ 
Norm$_{disk}$ & $980^{+420}_{-288}$& $950^{+370}_{-360}$ & $900^{+300}_{-240}$\\ 
kT$_{seed}$ (mean) & $1.09\pm0.06$ & $1.09^{+0.08}_{-0.06}$ & $1.13^{+0.07}_{-0.06}$\\ 
kT$_{seed}$(cov) & $1.65\pm0.05$ & $1.62\pm0.04$ &  $1.54\pm{0.1}$\\
kT$_{e }$ (tied) & $3.3\pm{+0.13}$ & $3.27^{+0.37}_{-0.26}$ & $3.4^{+0.5}_{-0.3}$\\ 
Norm$_{nthcomp}$(mean)& $(6.7^{+0.9}_{-1.2})\times10^{-2}$& $(6.7^{+0.5}_{-0.6})\times10^{-2}$ & $(7.0\pm{0.1})\times10^{-2}$\\ 
Norm$_{nthcomp}$(cov)& $(2.0\pm{0.2})\times10^{-3}$ & $(2.01\pm{0.09})\times10^{-3}$ & $(1.7\pm{0.3})\times10^{-3}$ \\ 
$\Gamma$ (tied) & $2.99\pm0.2$ & $2.97\pm{0.02}$ & $3.1^{+0.1}_{-0.2}$\\
E$_{line}$(tied) & $6.54^{+0.08}_{-0.1}$ & $6.53^{+0.09}_{-6.53}$ & $6.55\pm{0.05}$\\ 
$\sigma_{line}$(tied) & $0.69^{+0.15}_{-0.14}$ & $0.70^{+0.19}_{-0.13}$ & $0.49^{+0.04}_{-0.09}$\\ 
Norm$_{E_{line}}$(mean) & $(2.0^{+1.0}_{-0.4})\times10^{-3}$& $(2.2\pm{0.4})\times10^{-3}$ & $(1.6^{+0.4}_{-0.3})\times10^{-3}$ \\ 
Norm$_{E_{line}}$(cov) & $(4.8\pm{3})\times10^{-4}$& 0.0 (fixed) & 0.0 (fixed)\\ 
$\chi^{2}_{\nu}$ & 1.59 (44) & 1.78 (45) & 1.56 (35)
\enddata
\tablecomments{Obs. group 1* includes a Gaussian in modeling the covariance spectrum.  All other fits have no Gaussian in the covariance spectrum model.  All energies are given in keV.}
\label{tab:fit}	
\end{deluxetable*}

\section{Discussion} \label{discussion}
We have analyzed all {\it RXTE} data of Aql-X1 that show significant kHz QPOs and that were in modes with adequate resolution in time (< 128 $\mu s$) and  energy (64 channels).  This work was motivated by the desire to expand the scope of spectral--timing analysis of kHz QPOs to a wider array of neutron star LMXB systems.  We only analyzed the lower kHz QPO of Aql-X1 due to the poor S/N of the upper kHz -- which was only discovered in \citet{Barret_08}.  All analyses are associated with the lower kHz QPO.  As in \citet{Barret_13, deAvellar_13, Peille_15}, for objects 4U 1608-52, 4U 1636-53, and 4U 1728-34 respectively, we found soft lags between the high energy X-ray photons and low energy X-ray photons. The magnitude of lags in Aql X-1 were on the order of 30 $\mu s$ and comparable to all the previous studies of neutron star LMXB systems.  Additionally, over the QPO frequencies, we find large dependencies of lag on frequency are excluded, consistent with \citet{deAvellar_13, Peille_15}.  We note that \citet{Barret_13} does find some variation of lag with QPO frequency, since that work used a larger data set for 4U 1608-52 than \citet{deAvellar_13}.  See \citet{Barret_13} for a discussion of the magnitude of the average lag and its implication on the geometry of neutron star systems.

The shape and magnitude of the lag-energy spectra for Aql-X1 is also consistent with the other objects previously mentioned.  This includes a smooth decrease in the lags toward higher energies.  The exact mechanism and source of lags is poorly understood.  One possibility is that thermal Comptonization in the boundary layer causes the lags.  See \citet{Lee_01,Kumar_Misra_14,Kumar_Misra_16} for a discussion of different models of Comptonization and how they produce lags.  Another possible explanation of the production of lags is X-ray reflection.  In the reflection scenario, soft lags are thought to be associated with reverberation.  Here, a hard source of photons -- possibly the neutron star boundary layer formed at the point where the faster Keplerian motion of the accretion flow encounters the slower rotating neutron star surface -- impinges on and is reprocessed by the accretion disk.  Whereas hard lags are thought to arise due inward propagating accretion rate variations which modulate the hard Comptonized flux via seed photon fluctuations.  Additionally, lags can also be due to intrinstic, coherent spectral softening \citep{Kaaret_99} or due to temperature oscillations between two different non-isothermal Comptonizing sources \citep[e.g.,][]{deAvellar_13, Peille_15} which might indicate a composite Comptonizing source.

 \citet{Peille_15} point out that because the lag-energy spectrum drops at energies where the accretion disk does not contribute a significant amount of flux, there must be some property associated with Comptonization alone that must contribute to the lags.  Also, relative rms (covariance) increases above energies where the accretion disk should contribute to the flux and therefore the variations there are likely modulated by a harder source of photons, possibly the boundary layer \citep[see e.g.,][]{deAvellar_13}.  Recently, \citet{Cackett_16} modeled the lag-energy spectrum of 4U 1608-52 in order to test whether reverberation could produce the observed lags.  While finding that reverberation could account for the lags below 8 keV, the behavior of the lags above 8 keV was markedly different than predicted.  

Our spectral fits of the mean and covariance spectra in Aql X-1 yield similar results as \citet{Peille_15}.  We find systematically higher seed photon temperatures for the covariance spectra over the mean spectra.   Additionally, the covariance spectra are harder than the mean spectra, a result seen in all neutron star LMXBs to date and is well fit by a thermal Comptonized component \citep{Gilf_03,Peille_15}.  The implications of these findings are discussed in detail in \citet{Peille_15}.  

Finally, we have discovered that in one set of observations, the covariance spectrum is better fit with a combination of a thermal Comptonized component and a Fe K line Gaussian profile.  This hints at the possibility of a reflection/reverberation signature that contributes to the lags, at least in part, or that some other mechanism can modulate the Fe K line at the frequency of the lower kHz QPO.  
Interestingly, by taking the ratio or the iron line normalization in the time averaged and covariance spectra, the fractional RMS is $\simeq 24\%$ which is much higher than the observed fractional RMS which never exceeds $\simeq 10\%$ in this component of the spectrum; see Figure \ref{fig:rel_rms}.  This implies that Fe K line is more variable at the QPO frequency than the overall hard emission.  We do not have a physical explanation of this.

Currently, there are no models that explain all the spectral-timing properties of neutron star LMXBs.     

\section{Conclusion} \label{conclusion}
We have studied the spectral-timing properties of the neutron star LMXB Aql X-1.  We found similar behavior in the lag-frequency and lag-energy relationships as well as covariance spectral decompositions as seen previously in other neutron star LMXBs that have been studied.  This adds an additional source to those where detailed spectral-timing analysis of kHz QPOs has been done, and provides further support for the conclusions reached in all cases.  Specifically, the covariance spectra is well fit by a thermal Comptonized component and spectral fits indicate a higher seed photon temperature for the covariance spectrum.  This implies a possible composite boundary layer emitting region. 

We also find for one set of observations, the covariance spectrum is fit better with a thermal Comptonized component and Fe K line with 2.4-$\sigma$ confidence.  The implications of this are less clear.  While tempting to attribute this to reverberation, more information is needed.  Moreover, neither 4U 1608-52, nor 4U 1728-34 show this feature in their respective covariance spectra.  Spectral-timing analysis of additional sources is needed to determine if this result is more common in neutron star LMXBs.  Also, future missions with better spectral resolution -- while maintaining the high timing capability of {\it RXTE} -- might unlock this feature and help answer questions about the fundamental nature of accretion and emission of these objects.

\acknowledgements
JST and EMC gratefully acknowledge support from the National Science Foundation through CAREER award number AST-1351222.  The authors would like to thank Didier Barret, Philippe Peille and the participants of the Lorentz Center workshop `The X-ray Spectral-Timing Revolution' (February 2016), for useful discussions.  The authors also thank the referee for his careful review and comments which contributed significantly to the quality of this work.

\bibliographystyle{apj}
\bibliography{apj-jour,LMXB}

\begin{thebibliography}{32}
\expandafter\ifx\csname natexlab\endcsname\relax\def\natexlab#1{#1}\fi

\bibitem[{{Arnaud}(1996)}]{Arnaud_96}
{Arnaud}, K.~A. 1996, in Astronomical Society of the Pacific Conference Series,
  Vol. 101, Astronomical Data Analysis Software and Systems V, ed. G.~H.
  {Jacoby} \& J.~{Barnes}, 17

\bibitem[{{Barret}(2013)}]{Barret_13}
{Barret}, D. 2013, \apj, 770, 9

\bibitem[{{Barret} {et~al.}(2008){Barret}, {Boutelier}, \&
  {Miller}}]{Barret_08}
{Barret}, D., {Boutelier}, M., \& {Miller}, M.~C. 2008, \mnras, 384, 1519

\bibitem[{{Bradt} {et~al.}(1993){Bradt}, {Rothschild}, \& {Swank}}]{Bradt_93}
{Bradt}, H.~V., {Rothschild}, R.~E., \& {Swank}, J.~H. 1993, \aaps, 97, 355

\bibitem[{{Cackett}(2016)}]{Cackett_16}
{Cackett}, E.~M. 2016, \apj, 826, 103

\bibitem[{{de Avellar} {et~al.}(2016){de Avellar}, {M{\'e}ndez}, {Altamirano},
  {Sanna}, \& {Zhang}}]{deAvellar16}
{de Avellar}, M.~G.~B., {M{\'e}ndez}, M., {Altamirano}, D., {Sanna}, A., \&
  {Zhang}, G. 2016, \mnras, 461, 79

\bibitem[{{de Avellar} {et~al.}(2013){de Avellar}, {M{\'e}ndez}, {Sanna}, \&
  {Horvath}}]{deAvellar_13}
{de Avellar}, M.~G.~B., {M{\'e}ndez}, M., {Sanna}, A., \& {Horvath}, J.~E.
  2013, \mnras, 433, 3453

\bibitem[{{Gilfanov} {et~al.}(2003){Gilfanov}, {Revnivtsev}, \&
  {Molkov}}]{Gilf_03}
{Gilfanov}, M., {Revnivtsev}, M., \& {Molkov}, S. 2003, \aap, 410, 217

\bibitem[{{Kaaret} {et~al.}(1999){Kaaret}, {Piraino}, {Ford}, \&
  {Santangelo}}]{Kaaret_99}
{Kaaret}, P., {Piraino}, S., {Ford}, E.~C., \& {Santangelo}, A. 1999, \apjl,
  514, L31

\bibitem[{{Kalberla} {et~al.}(2005){Kalberla}, {Burton}, {Hartmann}, {Arnal},
  {Bajaja}, {Morras}, \& {P{\"o}ppel}}]{Kalberla_05}
{Kalberla}, P.~M.~W., {Burton}, W.~B., {Hartmann}, D., {Arnal}, E.~M.,
  {Bajaja}, E., {Morras}, R., \& {P{\"o}ppel}, W.~G.~L. 2005, \aap, 440, 775

\bibitem[{{Kumar} \& {Misra}(2014)}]{Kumar_Misra_14}
{Kumar}, N., \& {Misra}, R. 2014, \mnras, 445, 2818

\bibitem[{{Kumar} \& {Misra}(2016)}]{Kumar_Misra_16}
---. 2016, ArXiv e-prints

\bibitem[{{Lamb} \& {Miller}(2001)}]{lamb01}
{Lamb}, F.~K., \& {Miller}, M.~C. 2001, \apj, 554, 1210

\bibitem[{{Lee} {et~al.}(2001){Lee}, {Misra}, \& {Taam}}]{Lee_01}
{Lee}, H.~C., {Misra}, R., \& {Taam}, R.~E. 2001, \apjl, 549, L229

\bibitem[{{Lin} {et~al.}(2007){Lin}, {Remillard}, \& {Homan}}]{Lin_07}
{Lin}, D., {Remillard}, R.~A., \& {Homan}, J. 2007, \apj, 667, 1073

\bibitem[{{M{\'e}ndez} {et~al.}(2001){M{\'e}ndez}, {van der Klis}, \&
  {Ford}}]{Mendez_01}
{M{\'e}ndez}, M., {van der Klis}, M., \& {Ford}, E.~C. 2001, \apj, 561, 1016

\bibitem[{{M{\'e}ndez} {et~al.}(1998){M{\'e}ndez}, {van der Klis}, {van
  Paradijs}, {Lewin}, {Vaughan}, {Kuulkers}, {Zhang}, {Lamb}, \&
  {Psaltis}}]{Mendez_98}
{M{\'e}ndez}, M. {et~al.} 1998, \apjl, 494, L65

\bibitem[{{Miller} {et~al.}(1998){Miller}, {Lamb}, \& {Psaltis}}]{miller98}
{Miller}, M.~C., {Lamb}, F.~K., \& {Psaltis}, D. 1998, \apj, 508, 791

\bibitem[{{Nowak} {et~al.}(1999){Nowak}, {Vaughan}, {Wilms}, {Dove}, \&
  {Begelman}}]{Nowak_99}
{Nowak}, M.~A., {Vaughan}, B.~A., {Wilms}, J., {Dove}, J.~B., \& {Begelman},
  M.~C. 1999, \apj, 510, 874

\bibitem[{{Peille} {et~al.}(2015){Peille}, {Barret}, \& {Uttley}}]{Peille_15}
{Peille}, P., {Barret}, D., \& {Uttley}, P. 2015, \apj, 811, 109

\bibitem[{{Stella} \& {Vietri}(1999)}]{stella99}
{Stella}, L., \& {Vietri}, M. 1999, Physical Review Letters, 82, 17

\bibitem[{{Uttley} {et~al.}(2014){Uttley}, {Cackett}, {Fabian}, {Kara}, \&
  {Wilkins}}]{Uttley_14}
{Uttley}, P., {Cackett}, E.~M., {Fabian}, A.~C., {Kara}, E., \& {Wilkins},
  D.~R. 2014, \aapr, 22, 72

\bibitem[{{Uttley} {et~al.}(2011){Uttley}, {Wilkinson}, {Cassatella}, {Wilms},
  {Pottschmidt}, {Hanke}, \& {B{\"o}ck}}]{Uttley_11}
{Uttley}, P., {Wilkinson}, T., {Cassatella}, P., {Wilms}, J., {Pottschmidt},
  K., {Hanke}, M., \& {B{\"o}ck}, M. 2011, \mnras, 414, L60

\bibitem[{{van der Klis}(1998)}]{vanderKlis_98}
{van der Klis}, M. 1998, Advances in Space Research, 22, 925

\bibitem[{{van der Klis}(2000)}]{vanderKlis_00}
---. 2000, \araa, 38, 717

\bibitem[{{van der Klis}(2006)}]{vanderKlis_06}
---. 2006, Advances in Space Research, 38, 2675

\bibitem[{{van der Klis} {et~al.}(1987){van der Klis}, {Hasinger}, {Stella},
  {Langmeier}, {van Paradijs}, \& {Lewin}}]{vanderKlis_87}
{van der Klis}, M., {Hasinger}, G., {Stella}, L., {Langmeier}, A., {van
  Paradijs}, J., \& {Lewin}, W.~H.~G. 1987, \apjl, 319, L13

\bibitem[{{Vaughan} {et~al.}(1998){Vaughan}, {van der Klis}, {M{\'e}ndez}, {van
  Paradijs}, {Wijnands}, {Lewin}, {Lamb}, {Psaltis}, {Kuulkers}, \&
  {Oosterbroek}}]{Vaughan_98}
{Vaughan}, B.~A. {et~al.} 1998, \apjl, 509, L145

\bibitem[{{Wagoner}(2003)}]{Wagoner_03}
{Wagoner}, R.~V. 2003, \nat, 424, 27

\bibitem[{{Wilkinson} \& {Uttley}(2009)}]{Wilk_09}
{Wilkinson}, T., \& {Uttley}, P. 2009, \mnras, 397, 666

\bibitem[{{Zdziarski} {et~al.}(1996){Zdziarski}, {Johnson}, \&
  {Magdziarz}}]{Zdz_96}
{Zdziarski}, A.~A., {Johnson}, W.~N., \& {Magdziarz}, P. 1996, \mnras, 283, 193

\bibitem[{{{\.Z}ycki} {et~al.}(1999){{\.Z}ycki}, {Done}, \& {Smith}}]{Zycki_99}
{{\.Z}ycki}, P.~T., {Done}, C., \& {Smith}, D.~A. 1999, \mnras, 309, 561

\end{thebibliography}
\end{document}